\documentclass[aoas]{imsart}

\usepackage{amsthm,amsmath}
\usepackage[authoryear]{natbib}
\RequirePackage[colorlinks,citecolor=blue,urlcolor=blue]{hyperref}
\RequirePackage{hypernat}

\usepackage{bm }
\usepackage{graphicx}
\usepackage{threeparttable}

\usepackage{color}
\startlocaldefs

\newtheorem{thm}{Theorem}[section]

\newtheorem{lem}[thm]{Lemma}

\newtheorem{rem}[thm]{Remark}

\setattribute{tablecaption}{shape}{}

\setattribute{tablename} {skip}{.~}
\newcommand{\norm}[1]{\left\Vert#1\right\Vert}

\newcommand{\R}{\mathbb R}
\newcommand{\eps}{\varepsilon}

\newcommand{\ds}{\mathrm{d}s}

\newcommand{\Cov}{\textrm{Cov}}

\newcommand{\bvec}{\bm{b}}
\newcommand{\bveca}{\bm{b}_{\bm{1}}}
\newcommand{\bvecb}{\bm{b}_{\bm{2}}}
\newcommand{\bveci}{\bm{b}_{\bm{i}}}
\newcommand{\bvecj}{\bm{b}_{\bm{j}}}

\newcommand{\fvec}{\bm{f}}
\newcommand{\phivec}{\bm{ \phi} }
\newcommand{\psivec}{\bm{ \psi} }

\newcommand{\muvec}{\bm{\mu}}
\newcommand{\muveci}{\bm{\mu}_{\bm{i}}}
\newcommand{\muveca}{\bm{\mu}_{\bm{1}}}
\newcommand{\muvecb}{\bm{\mu}_{\bm{2}}}

\newcommand{\xvec}{\bm{x}}
\newcommand{\xveci}{\bm{x}_{\bm{i}}}
\newcommand{\xveca}{\bm{x}_{\bm{1}}}
\newcommand{\xvecb}{\bm{x}_{\bm{2}}}

\newcommand{\yveci}{\bm{y}_{\bm{i}}}
\newcommand{\yveca}{\bm{y}_{\bm{1}}}

\newcommand{\Xa}{X_{1}}
\newcommand{\Xb}{X_{2}}
\newcommand{\Xhat}{\hat{X}_{2}}
\newcommand{\Xtilde}{\tilde{X}_{2}}
\newcommand{\Ya}{Y_{1}}
\newcommand{\Yb}{Y_{2}}

\newcommand{\Gammai}{\Gamma_{11}^+}
\newcommand{\Gai}{G_{11}^+}

\newcommand{\hvec}{\bm{h}}

\newcommand{\epsvec}{\bm{ \epsilon}}

\newcommand{\Var}{\textrm{Var}}

\endlocaldefs

\begin{document}

\begin{frontmatter}

\title{The Best Linear Unbiased Estimator for Continuation of a Function}
\runtitle{The BLUP for Continuation of a Function}


\author{\fnms{Yair} \snm{Goldberg}\thanksref{m1}\ead[label=e1]{yair.goldberg@mail.huji.ac.il}},
\author{\fnms{Ya'acov} \snm{Ritov}\thanksref{m1}\ead[label=e2]{yaacov.ritov@huji.ac.il}}
\and
\author{\fnms{Avishai} \snm{Mandelbaum}\thanksref{m2}\ead[label=e3]{avim@tx.technion.ac.il}}

\affiliation{The Hebrew University\thanksmark{m1} and Technion-—Israel Institute of Technology\thanksmark{m2}}

\address{Yair Goldberg and Ya'acov Ritov\\Department of Statistics and\\
The Center for the Study of Rationality\\
The Hebrew University \\
Jerusalem 91905, Israel\\
\printead{e1}\\
\phantom{E-mail:\ }\printead*{e2}}

\address{Avishai Mandelbaum\\Industrial Engineering and Management\\
Technion—-Israel Institute of Technology\\ Haifa 32000, Israel\\
\printead{e3}}
\runauthor{Goldberg et al.}

\begin{abstract}

We show how to construct the best linear unbiased predictor (BLUP)
for the continuation of a curve in a spline-function model. We
assume that the entire curve is drawn from some smooth random
process and that the curve is given up to some cut point. We
demonstrate how to compute the BLUP efficiently. Confidence bands
for the BLUP are discussed. Finally, we apply the proposed BLUP to
real-world call center data. Specifically, we forecast the
continuation of both the call arrival counts and the workload
process at the call center of a commercial bank.
\end{abstract}


\end{frontmatter}

\section{Introduction}\label{sec:intro}
Many data sets consist of a finite number of
multidimensional observations, where each of these
observations is sampled from some underlying smoothed
curve. In such cases it can be advantageous to address the
observations as functional data rather than as multiple
series of data points. This approach was found useful, for
example, in noise reduction, missing data handling, and in
producing robust estimations \citep[see the books][for a
comprehensive treatment of functional data
analysis]{bookRamsayApplied,bookRamsay}. In this work we
consider the problem of forecasting the continuation of a
curve using functional data techniques.


The problem we consider here is relevant to longitudinal
data sets, in which each observation consists of a series
of measurements over time that describe an underlying
curve. Examples of such curves are growth curves of
different individuals and arrival rates of calls to a call
center or of patients to an emergency room during different
days. We assume that such curves, or measurement series
that approximate these curves, were collected previously.
We would like to estimate the continuation of a new curve
given its beginning, using the behavior of the previously
collected curves.


Although each observation consists of a finite number of
points, the observation can be thought of as a smooth
function. This dual representation leads to two different
approaches when attempting to solve the prediction problem.
In the discrete approach, each observation is a
longitudinal vector of length $p+q$. We are interested in
the prediction of the last $q$-length part of the new
observation, given its beginning $p$-length part. This can
be computed by treating the beginning $p$-length vector as
the predictor variables and the last $q$-length vector as
the response variables. A prediction can be found, for
example, by finding the best linear unbiased predictor (see
\eqref{eq:multivariate_estimator}). The disadvantage of the
discrete approach is that the smooth nature of the
underlying function is ignored. If, instead, the continuous
approach is used, the prediction problem might  be treated
naively using regression techniques in which both the
predictor and the response are functions \citep[][Chapter
16]{bookRamsay}. However, these techniques do not take into
account the fact that the response function is a smooth
continuation of the predictor function.


In this paper, we choose the continuous approach.
Specifically, we would like to generalize the discrete case
to the continuous one, taking the smooth nature of the
curves into account. There are three main points that need
to be addressed. First, the curves lie within an infinite
dimensional space, while the number of observed curves is
finite. This indicates that a simple model for description
of the data is required. Second, the full-length curves,
the curve beginnings, and the curve continuations all lie
in different functional spaces, which, in contrast to the
discrete case, cannot generally be related by a linear
projection. Third, we require that the prediction should be
a smooth continuation of the beginning of the curve (at
least in the absence of noise).

Forecasting of the continuation of a function was
considered in previous works.
\citet{BesseCardotStephenson_Autoregressive2000}, and
\citet{Antoniadis_autoregressive2006}, among others,
developed different techniques for curve-valued
autoregressive processes. In these models, each curve
describes some longitudinal data cycle such as climate
variation during a year
\citep{BesseCardotStephenson_Autoregressive2000} and
television audience rates during a day
\citep{Antoniadis_autoregressive2006}. These models assume
that the cycles behave according to some autoregressive
model. The aim of these works is to predict the next cycles
given past observed cycles. The continuity point at the
beginning of each cycle, if it exists, is usually not taken
into account. The model discussed in \citet{Shen_2009} is
closer to ours. \citeauthor{Shen_2009} discusses a
curved-valued time series model in which past curves were
previously observed, and the beginning part of a new curve
is given. \citeauthor{Shen_2009} first forecasts the new
curve entirely, and then updates this forecast based on the
given curve beginning using penalized least squares.
However, all the models discussed above assume some time
series behavior, while the model discussed here assumes
that the curve-valued observations are independent.


The forecasting of curve continuation suggested here is
based on finding the best linear unbiased predictor (BLUP)
\citep{BLUP_Robinson91}. We assume that the curves are
governed by a small number of factors, possibly with
additional noise. These factors determine the main
variation between the different curves. The computation of
the predictor is performed in two steps. First, the
factors' coefficients are estimated from the beginning of
the new curve, which is defined on the first part of the
segment. Second, the prediction is obtained by computing
the representation of the factors on the latter part of the
segment. We prove that the resulting estimator is indeed
the BLUP and that it is a smooth continuation of the
beginning of the curve (at least in the absence of noise).

The two-step procedure for obtaining the BLUP involves
computation of the mean function on both partial segments,
and of the covariance operator on both segments and between
them, which can be computationally demanding if not
performed prudently. We approximate the curve data using a
spline function space of (possibly large) finite-dimension
\citep{bookDeBoor}. More specifically, we represent the
curves using appropriate B-splines bases. The use of
splines is common in functional data analysis due to the
simplicity of spline computation, and the ability of
splines to approximate smooth functions. We take advantage
of two more attributes of finite-dimensional spline
functional spaces. First, the functional space restriction
from the whole segment to a partial segment (the beginning
part or the latter part) has a natural B-spline basis that
has a lower number of elements. This solves collinearity
problems which can render any projection on the partial
segment basis instable. Second, the knot-insertion
algorithm \citep[see][Chapter 11]{bookDeBoor} ensures an
efficient and stable way to compute the mean function and
covariance operators on different partial segments.

The proposed forecasting procedure yields a smooth curve
which is the best linear unbiased prediction. Note,
however, that the continuation part of the function is
random, and therefore requires confidence bands. We present
confidence bands for the prediction, following
\citet{KnaflSacksYlvisaker_band1985}, under the assumption
that the curves aries from a Gaussian process. The bands
are computed in two steps. First, confidence intervals are
computed simultaneously for a finite set of points. Then,
using the fact that splines are piecewise polynomials, a
global band is found. We also suggest a way to compute
confidence bands using cross-validation. While no
theoretical justification proof is given for the cross
validation  confidence bands, they are much faster to
compute, and the numerical examples in
Section~\ref{sec:numerical} show that this approach works
considerably well.

We apply the forecasting procedure suggested here to call
center data. We forecast the continuation of two processes:
the arrival process and the workload process \citep[i.e.,
the amount of work in the system; see, for
example,][]{Aldor_2009}. In call centers, the forecast of
the arrival process plays an important roll in determining
staffing levels. Optimization of the latter is important
since salaries account for about 60-70\% of the cost of
running a call center \citep{Gans2003}. Usually, call
center managers utilize forecasts of the arrival process
and knowledge of service times, along with some
understanding of customer patience characteristics
\citep{Zeltyn_2005}, to estimate future workload and
determine staffing level \citep{Aldor_2009}. The
disadvantage of this approach is that the forecast of the
workload is not performed directly, and instead it is
obtained using the forecast of the arrival process.
\citet{Michael_Thesis} showed how the workload process can
be estimated explicitly, thereby enabling direct forecast
of the workload. In this work we forecast the continuation
of both the arrival and workload processes, given past
days' information and the information up to some time of
the day. We compare between the results for the arrival
process and the workload process. We also compare our
results for the arrival process to those of other
forecasting techniques, namely, to the techniques that were
introduced by \citet{Weinberg2007} and
\citet{ShenHuang_2008}.

The paper is organized as follows. The functional model and
notation are presented in Section~\ref{sec:notation}. The
main theoretical results are presented in
Section~\ref{sec:bestPredictor}, were we first show how to
construct the BLUP for the continuation of a curve. Next,
we show how the BLUP can be computed efficiently.
Confidence bands are discussed in
Section~\ref{sec:confidence}. In
Section~\ref{sec:numerical} we apply the estimator to
real-world data, comparing direct and indirect workload
forecasting, and comparing our results to other techniques.
Concluding remarks appear in Section~\ref{sec:summary}.
Technical proofs are provided in the Appendix.
\section{The functional model}\label{sec:notation}
In this section we present the model and notation that will
be used throughout this paper. Let $X$ be a random function
defined on the segment $S=[0,T]$, and let the random
functions $\Xa$ and $\Xb$ be the restrictions of $X$ to the
segments $S_1=[0,U]$ and $S_2=[U,T]$, respectively, for
some $0< U<T$. Our goal is to estimate $X_2$ given
information regarding $X_1$.

We assume that $X$ is of the form
\begin{equation*}
    X(t)=\mu(t)+\phivec(t)'\hvec\,,
\end{equation*}
where $\mu(t)$ is the mean function,
$\hvec=(h_1,\ldots,h_p)$ is a random vector with mean zero
and covariance matrix $L$, $L$ is diagonal with $L_{11}\geq
\ldots\geq L_{pp} >0$, and
$\phivec(t)=(\phi_1(t),\ldots,\phi_p(t))'$ is a vector of
orthonormal functions. We assume that the functions $\mu$
and $\phi_j$ have a basis expansion with respect to some
B-spline basis $\bvec=(b_1,...,b_N)'$, defined on some
fixed knot sequence $\tau$. We denote this B-spline space
by $S_{k,\tau}$ where $k$ denotes the the splines' order.
Thus, we can write $\mu(t)=\bvec(t)'\muvec$ and
$\phivec(t)= A '\bvec(t) $, for some $p\times 1$ vector
$\muvec$ and $N\times p$ loading matrix $A$. Thus, we have
\begin{equation}\label{not:vecX}
    X(t)=\bvec(t)'\big(\muvec +A \hvec\big)\doteq\bvec(t)'\xvec\,,
\end{equation}
where $\xvec=\muvec +A \hvec$. We think of  $N$, the
ambient functional space dimension, as being much larger
then $p$, the dimension of the subspace which spanned by
the random function $X$.

We assume that instead of seeing $X$, we actually observe
some noisy version of $X$, namely
\begin{equation*}
    Y(t)=X(t)+\eps(t)\,,
\end{equation*}
where $\eps(t)=\psivec(t)'\epsvec$ is some random function
independent of $X(t)$, $\epsvec$ is a $q\times 1$ zero-mean
random vector with diagonal covariance matrix $\Sigma$, and
$\psivec$ is a vector of functions. Since $X(t)$ is a
(random) linear combination of
$\phi_1(t),\ldots,\phi_p(t)$, we consider the noise as the
part of the observed function $Y(t)$ that cannot be
explained using such linear combinations. Hence we assume
that $\psivec$ is orthogonal to $\phivec$. However, note
that this orthogonality is not necessarily preserved when
$\psivec$ and $\phivec$ are restricted to one of the
segments $S_1$ or $S_2$. We assume that $\psivec$ also has
an expansion with respect to the basis $\bvec$ and hence
$\psivec(t)=B'\bvec(t)$ for some $N\times q$ loading matrix
$B$. Using this notation we may write
\begin{equation}\label{not:vecY}
    Y(t)=\bvec(t)'\big(\muvec +A \hvec+B\epsvec\big)\,.
\end{equation}

The covariance functions $u(s,t)=\Cov(X(s),X(t))$ and
$v(s,t)=\Cov(Y(s),Y(t))$ can be written by $\bvec(s)'
(ALA')\bvec(t)\doteq \bvec(s)' g \bvec(t)$ and $\bvec(s)'
(ALA'+B\Sigma B')\bvec(t)\doteq \bvec(s)' G \bvec(t)$,
respectively. We define the correspondence covariance
operators from $S_{k,\tau}$ to itself for functions $f\in
S_{k,\tau}$ as
\begin{eqnarray*}
        (\gamma f)(t)&=&\int_S u(s,t)f(s)\ds=\bvec(t)'g W \fvec\\
    (\Gamma f)(t)&=&\int_S v(s,t)f(s)\ds=\bvec(t)'G W \fvec
\end{eqnarray*}
where $W=\int_S \bvec(s)\bvec(s)'\ds$, and $\fvec$ is the
expansion of the function $f$ in the B-spline basis.

We now introduce the notation for $\Xa$ and $\Xb$ and their
respective noisy versions $\Ya$ and $\Yb$. Let $\tau_1$ and
$\tau_2$ be knot sequences that agree with $\tau$ on the
segments $[0,U)$ and $(U,T]$, respectively, and have knot
multiplicity of $k$ at $U$. Let $S_{k,\tau_i}$ for $i=1,2$
be the $k$-ordered spline space with knot sequence $\tau_i$
and let $\bveci(t)=(b_{i1}(t),\ldots,b_{iN_i}(t))$ be its
corresponding B-spline basis. We wish to represent $X_i$
and $Y_i$ ($i=1,2$) using the representations of $X$ and
$Y$.

Note that when the functions $\mu(t)$, $\phi_j(t)$,
$\psi_j(t)$, $v(s,t)$ and $u(s,t)$ are known on $[0,T]$,
they are also known on $S_1$ and $S_2$. Thus, it is enough
to represent these functions using the bases $\bveci$ in
order to obtain representations for $X_i$ and $Y_i$. Recall
that $\mu(t)=\bvec(t)'\muvec$ for some vector of
coefficients $\muvec$. Using the knot-insertion algorithm
\citep[see][Chapter 11]{bookDeBoor} we obtain new vectors
$\muveci$ such that (a) $\mu(t)=\bveci(t)'\muveci$ for all
$t$ on which $\bveci$ is defined and (b) $\muveci$ is
obtained from $\muvec$ by truncation and a change of at
most $k$ coefficients. Similarly, using the knot-insertion
algorithm, we can obtain the loading matrices $A_i$ and
$B_i$ such that $\phivec(t)=A_i \bveci(t)$ and
$\psivec(t)=B_i \bveci(t)$ for all $t$ on which $\bveci$ is
defined. Summarizing, we have
\begin{eqnarray}\label{not:vecXi}
    X_i(s)&=&\bveci(s)'\big(\muveci +A_i \hvec\big)\doteq\ \bveci(s)'    \xveci\\
        Y_i(s)&=&\bveci(s)'\big(\muveci +A_i \hvec+B_i\epsvec\big)\doteq \bveci(s)'    \yveci\nonumber\\
        v(s,t)&=& \bveci(s)(A_i L A_j ' +B_i \Sigma B_j')\bvecj(t)\doteq
\bveci(s)'G_{ij}\bvecj(t)\nonumber\\
u(s,t)&=& \bveci(s)(A_i L A_j ')\bvecj(t)\doteq
\bveci(s)'g_{ij}\bvecj(t)\nonumber
\end{eqnarray}
for $i,j=1,2$ and for each $s\in S_i$ and $t\in S_j$.

We define the operators $\gamma_{ij}$ and $\Gamma_{ij}$
from $S_{k,\tau_j}$ to $S_{k,\tau_i}$ for $i,j=1,2$ by
\begin{eqnarray}\label{not:Gamma_ij}
    (\gamma_{ij} f)(t)&=&\int_{S_j} u(s,t)f(s)\ds=\bveci(t)'g_{ij} W_j \fvec\\
       (\Gamma_{ij} f)(t)&=&\int_{S_j} v(s,t)f(s)\ds=\bveci(t)'G_{ij} W_j \fvec\,,\nonumber
\end{eqnarray}
where $W_j=\int_{S_j} \bvecj(s)\bvecj(s)'\ds$, and $\fvec$
is the expansion of the function $f$ in $\bvecj$.

The model discussed above will be used for the estimation
of $X_2$ given $Y_1$. Note that the distributions of $X$
and $Y$ are generally not known. In a realistic situation
one needs to estimate the model components. Recall that
$Y(t)=\bvec(t)'\big(\muvec +A \hvec+B\epsvec\big)$, where
$\hvec$ and $\epsvec$ are random vectors with zero mean and
covariance matrices $L$ and $\Sigma$, respectively. Before
discussing the forecasting procedure, we briefly discuss
how estimation of $\muvec, L, \Sigma $ and the loading
matrices $A$ and $B$ can be performed.

Assume that the functions $Y^{(1)},\ldots,Y^{(m)}$ were
drawn according to the distribution law of $Y$. We
distinguish between two scenarios. In the first scenario we
assume that the functions $Y^{(1)},\ldots,Y^{(m)}$ were
observed. In this case one can estimate the various
components of $Y$ using functional principal component
analysis (functional PCA). This can be done either by using
PCA on the coefficients of the functions or by introducing
some smoothness using regularized functional PCA
\citep[see, for example,][Chapters~8 and~9]{bookRamsay}.
The matrices $L$ and $\Sigma$ are then determined by the
eigenvalues of the PCA decomposition while the loading
matrices $A$ and $B$ consist of the coefficients of the
principal components with respect to the basis $\bvec$. The
size of $L$ and $\Sigma$ can be estimated using the gaps in
the eigenvalues of the PCA decomposition.

In the second scenario, we assume that some noisy discrete
observations are given; for example in the following form
\begin{equation*}
    Z^{(i)}(t_{ij})=Y^{(i)}(t_{ij})+e_{ij}\,,
\end{equation*}
for $i=1,\ldots,m$, $j=1,\ldots,n_j$, and $0\leq
t_{i1}<\ldots<t_{in_j}\leq T$, and where $e_{ij}\sim
N(0,\sigma^2)$ are independent. In this case, one can first
estimate the functions and then use functional PCA as
described above. The simplest way to estimate the functions
is to estimate each function separately, using, for
example, regression splines
\citep[][Chapter~14]{bookDeBoor}. This method is used in
the numerical examples in Section~\ref{sec:numerical}.
Others, such as \citet{Kneip94} and
\citet{BesseCardotFerraty97}, suggest to estimate all the
functions  simultaneously. Both methods use some sort of
functional PCA. These methods suggest ways to estimate the
length of $\hvec$. The method by
\citet{BesseCardotFerraty97} also assumes a splines
environment, as in our case.

\section{The construction of the BLUP}\label{sec:bestPredictor}

Given $\Ya$, the noisy version of the beginning part of the
random function $X$, our goal is to find a \emph{good}
estimator for $\Xb$, the continuation of $\Xa$.

Following~\citet{BLUP_Robinson91}, we say that $\Xhat$ is a
\emph{good} estimator of $\Xb$ given $\Ya$ if the following
criteria hold:
\begin{enumerate}\renewcommand{\labelenumi}{(C\arabic{enumi})}
  \item $\Xhat$ is a linear function of $\Ya$.
      \label{cr:linear}
  \item $\Xhat$ is unbiased, i.e.,
      $E[\Xhat(t)]=\mu(t)$.\label{cr:unbiased}
  \item $\Xhat$ has minimum mean square error among the
      class of linear unbiased
      estimators.\label{cr:MSE}
\end{enumerate}
Two more demands regarding the estimator that seems
desirable in our context are
\begin{enumerate}\renewcommand{\labelenumi}{(C\arabic{enumi})}
\addtocounter{enumi}{3}
\item The random function $\Xhat$ lies in the space
    $S_{k,\tau_2}$. \label{cr:continuous}
\item When no noise is introduced, i.e., when
    $\Ya=\Xa$, the concatenation of $\Xhat$ to $\Xa$
    lies in $S_{k,\tau}$; in other words, the combined
    function
  \begin{equation*}
    \hat{X}=\left\{
              \begin{array}{cc}
                \Xa(t) & 0\leq t\leq U \\
                \Xhat(t) & U< t \leq T \\
              \end{array}
            \right.
 \end{equation*} is smooth enough. \label{cr:concatenation}
\end{enumerate}
An estimator that fulfills
(C\ref{cr:linear})-(C\ref{cr:concatenation}) will be
referred to as a best linear unbiased predictor (BLUP). In
this section we will show how to  construct such a BLUP and
prove that is is defined uniquely.

\begin{rem}
Note that the definition of unbiased estimator in
(C\ref{cr:unbiased}) is not the usual definition. A more
restrictive criterion is
\begin{enumerate}\renewcommand{\labelenumi}{(C\arabic{enumi}*)}
\addtocounter{enumi}{1}
  \item $\Xhat$ is unbiased in the the following sense
      $E[\Xhat(t)|\Ya]=E[\Xb(t)|\Ya]$.\label{cr:unbiasedStrong}
\end{enumerate}
We will show that when $Y$ is a Gaussian process, this
criterion is fulfilled by the proposed BLUP as well.
\end{rem}

\begin{rem}The analogous results in the multivariate case are well known.
Here \emph{best} estimator means estimator that meets
criteria (C\ref{cr:linear})-(C\ref{cr:MSE}). Let
$Z=(Z_1,Z_2)'$ be a random vector such that
\begin{equation*}
     E\left[ \begin{array}{c} Z_1\\Z_2 \\ \end{array}\right]
    =\left( \begin{array}{c} m_1\\ m_2 \\
       \end{array}   \right)\,,\quad \mathrm{Var}\left[ \begin{array}{c}
            Z_1\\ Z_2 \\ \end{array} \right]=
            R =\left( \begin{array}{cc} R_{11} & R_{12} \\R_{21} & R_{22}\\
\end{array} \right)\,.
\end{equation*}
Then the BLUP of $Z_2 | Z_1$ is given by
\begin{equation}\label{eq:multivariate_estimator}
   \hat{Z}_2 =m_2+R_{21}R_{11}^{+}(Z_1-m_1)
\end{equation}
where $R_{11}^{+}$ is the Moore-Penrose pseudoinverse of
$R_{11}$ \citep[see, for
example,][]{singularCovariance_Marsaglia64}.
\end{rem}

In the following, we define the linear operators that are
the analogs of the matrices $R_{11}^{+}$ and $R_{2 1}$ from
the multivariate case. This enables the construction of a
uniquely-defined BLUP for $\Xb$.

We begin with defining the operator
$\Gammai:S_{k,\tau_1}\rightarrow S_{k,\tau_1} $, which is
the functional equivalent of $R_{11}^{+}$. Define the
function
\begin{equation*}
   v_{11}^{+}(s,t)= \bveca(s)' W_1^{-1}\Gai W_1^{-1}\bveca(t)\,,
\end{equation*}
for every $s,t\in S_1$. Note that $W_1$ is invertible since
it is a Gram matrix of basis functions \citep[see][Theorem
1.5]{Sansone59}. Define
\begin{equation*}
(\Gammai f)(t)=\int_{S_1}  v_{11}^{+}(s,t)f(s)\ds = \bveca(t)'
W_1^{-1}\Gai\fvec \,,
\end{equation*}
where $\fvec$ is the expansion of the function $f$ in the
B-spline basis $\bveca$. The following lemma justifies the
notation of $\Gammai$ as a pseudoinverse operator.
\begin{lem}\label{lem:psaudoinverseWorks}
With probability one,
\begin{equation*}
    \Gamma_{11} \Gammai (\Ya-\mu)
    =\Gammai\Gamma_{11}(\Ya-\mu)=\Ya-\mu\,.
\end{equation*}
\end{lem}
See proof in the Appendix.

We are now ready to define the estimator for $\Xb$ given
$\Ya$, similarly to
estimator~\eqref{eq:multivariate_estimator} in the
multivariate case, by
\begin{equation}\label{eq:Xhat}
\Xhat(t)=\mu(t)+\gamma_{21}
\Gammai(\Ya-\mu)(t)=\bvecb(t)'\big(\muvecb+ g_{21}  \Gai
(\yveca-\muveca) \big)\,,
\end{equation}
for every $t\in S_2$. Then we have
\begin{thm}\label{thm:hxisBLUP}
The estimator $\Xhat$ meets criteria
(C\ref{cr:linear})-(C\ref{cr:concatenation}) and is unique
up to equivalence. Moreover, if $Y$ is a Gaussian process,
then $\Xhat$ meets criterion (C\ref{cr:unbiasedStrong}*) as
well.
\end{thm}
\begin{proof}
We show that (C\ref{cr:linear})-(C\ref{cr:concatenation})
hold, one by one.

(C\ref{cr:linear}) holds because $\Xhat$ is indeed a linear
transformation of $\Ya$ as can be seen
from~\eqref{eq:Xhat}.

(C\ref{cr:unbiased}) holds since
\begin{equation*}
E[\Xhat(t)] = \bvecb(t)'\big(\muvecb+ g_{21} \Gai
(E[\yveca-\muveca]) \big)=\bvecb(t)'\muvecb=\mu(t)\,.
\end{equation*}

(C\ref{cr:MSE}) states that $\Xhat$ should minimize the
mean square error among all the unbiased linear estimators
. Let $\Xtilde$ be another linear unbiased estimator. Then
we can write $\Xtilde=(\Xtilde-\Xhat)+\Xhat$. Since both
$\Xtilde$ and $\Xhat$ are unbiased, $\Xtilde-\Xhat$ is an
unbiased linear estimator of zero, hence it is of the form
$\bvecb(t)' M(\yveca-\muveca) $ for some $N_2\times N_1$
matrix $M$. Moreover, it can be shown that
$\Cov(\Xb-\Xhat,\Xtilde-\Xhat)=0$. Indeed,
\begin{eqnarray*}
    \Cov\big((\Xb-\Xhat)(s),(\Xtilde-\Xhat)(t)\big)&=&E[(\Xb-\Xhat)(\Xtilde-\Xhat)(t)]\\
    &=& \bvecb(s)'
    E[(\xvecb-\muvecb)(\yveca-\muveca)']M'\bvecb(t)\\
    &&\;-\bvecb(s)' E[\muvecb+g_{21}\Gai(\yveca-\muveca)(\yveca-\muveca)']M'\bvecb(t)\\
     &=& \bvecb(s)'\big(g_{21}M'+ g_{21}\Gai
     G_{11}M')\big)\bvecb(t)=0\,.
\end{eqnarray*}
where the last equality follows from
Lemma~\ref{lem:psaudoinverseWorks}.

To see that $\Xhat$ minimizes the mean square error, note
that
\begin{eqnarray}\label{eq:hxIsThe Best}
  E[(\Xb-\Xtilde)^2(t)] &=& E[(\Xb-\Xhat)^2(t)]+
  E[(\Xtilde-\Xhat)^2(t)]+2 E[(\Xb-\Xhat)(\Xhat-\Xtilde)(t)] \nonumber\\
  &=& E[(\Xb-\Xhat)^2(t)]+E[(\Xtilde-\Xhat)^2(t)]\geq   E[(\Xb-\Xhat)^2(t)]\,,
\end{eqnarray}
which proves that $\Xhat$ minimizes the mean square error
and is unique up to equivalence.

(C\ref{cr:continuous}) holds by construction.

(C\ref{cr:concatenation}) states that when no noise is
introduced, $\Xhat$ a smooth continuation of $\Xa$. First,
note that by Lemma~\ref{lem:psaudoinverseWorks}
\begin{equation*}
 \Xa(t)=\bveca(t)'\big(\muveca+G_{11}\Gai(\xveca-\muveca)\big)=\bveca(t)'\big(\muveca+A_1 (L A_1 ' \Gai)(\xveca-\muveca)\big)\,.
\end{equation*}
By definition we also have
\begin{equation*}
 \Xhat(t)=\bvecb(t)'\big(\muvecb+g_{21}\Gai(\xveca-\muveca)\big)=\bvecb(t)'\big(\muvecb+A_2 (L A_1 ' \Gai)(\xveca-\muveca)\big)\,.
\end{equation*}
Define $\hat{X}(t)=\bvec(t)'\big(\mu(t)+A(L A_1 '
\Gai)(\xveca-\muveca)\big)$. It follows from the
definitions of $\muveci,A_i$ and $\bveci$ that $\hat{X}(t)$
agrees with $\Xa$ on $S_1$ and with $\Xhat$ on $S_2$. Since
$\hat{X}\in S_{k,\tau}$, the result follows.

Finally, if $Y$ is a Gaussian process, then $\yveca$ and
$\xvecb$ are normally distributed such that
$\Var(\yveca)=G_{11}$ and $ \Cov(\xvecb,\yveca)=g_{21}$.
Following~\cite{singularCovariance_Marsaglia64} we obtain
\begin{eqnarray}\label{eq:Gaussian}
  E[\Xb(t)|\Ya] &=& \bvec(t)'E[\xvecb|\yveca]=
    \bvec(t)'\big(\muvecb+g_{21}\Gai(\yveca-\muveca)\big) \\
   &=& \Xhat(t)=E[\Xhat(t)|\Ya]\nonumber
\end{eqnarray}
and criterion (C\ref{cr:unbiasedStrong}*) is met.
\end{proof}

It should be noted that when the parameters of the model
are estimated (see end of Section~\ref{sec:notation}) and a
Gaussian model is assumed, the estimator $\Xhat$ can be
considered as an empirical Bayes estimator. Indeed, the
estimation of the distribution of $\hvec$ and $\epsvec$ can
be considered as estimating the prior distribution, while
the the computation of $\Xhat$ as in~\eqref{eq:Gaussian} is
in fact finding the posterior mean given the data $\Ya$.

From a computational point of view, the computation of
$\Xhat$ may seem heavy. Indeed by~\eqref{eq:Xhat} it
involves finding the pseudoinverse of $G^{+}_{11}$ which is
an $N_1\times N_1$ matrix. However, a simpler expression
can be found. Recall that
\begin{equation*}
 G_{11}=[A_1,B_1]\left[  \begin{array}{cc}
  L&0\\0 &\Sigma
  \end{array} \right]\left[
           \begin{array}{c}
          A_1 '\\
          B_1 '   \\
           \end{array}
         \right]\doteq C S C' \,.
\end{equation*}
where $C=[A_1, B_1]$ and $S=\left[  \begin{array}{cc}
  L&0\\0 &\Sigma
  \end{array} \right]$. Using Lemma~\ref{lem:psaudo_inverse}.\ref{lem:psaudo_inverse3} with $T=S^{1/2} C'$ we have
\begin{eqnarray*}
  \Gai &=& CS^{1/2}\left(\left(S^{1/2}C'C S^{1/2}\right)^+ \right)^2 S^{1/2}C'\\
   &=& CS^{1/2}\left(S^{-1/2}(C'C)^+ S^{-1}(C'C)^+ S^{-1/2}\right) S^{1/2}C'\\
   &=& C(C'C)^+  S^{-1}(C'C)^+  C'\,,
\end{eqnarray*}
which involves the pseudoinverse computation of a $(p+q)
\times(p+q)$ matrix.

Finally, instead of assuming that
$\Ya(t)=\bveca(t)'\big(\muveca +A_1 \hvec+B_1\epsvec\big)$,
one may assume that
\begin{equation*}
   \Ya(t)=\bveca(t)'\big(\muveca+A_1 \hvec+ \bm{\tilde{\eps}}_{\bm{1}}\big)
\end{equation*}
where $\bm{\tilde{\eps}}_{\bm{1}}$ is a $N_1\times 1$ mean
zero random vector with $\sigma^2 I$ covariance matrix and
$I$ is the identity matrix. In this case,
\begin{equation}\label{eq:ridge}
    \Xhat(t)=\bvecb(t)'\big(\muvecb+g_{21}(A_1 L A_1'+\sigma^2 I)^{-1}(\xveca-\muveca)\big)
\end{equation}
which is the ridge regression estimator
\citep{HoelKennard70}. Once again, a simpler expression can
be obtained using some matrix algebra \citep[see][
Eq.~5.2]{BLUP_Robinson91}. We have
\begin{equation*}
    g_{21}(A_1 L A_1'+\sigma^2 I)^{-1}=A_2 L A_1 ' (A_1 L A_1'+\sigma^2
    I)^{-1}=A_2 \left(A_1'
    A_1+\sigma^2 L^{-1}\right)^{-1}A_1 '\,,
\end{equation*}
and hence $\Xhat(t)=\bvecb(t)'\big(\muvecb+A_2 \left(A_1'
A_1+\sigma^2 L^{-1}\right)^{-1}A_1 '(\xveca-\muveca)\big)$,
which involves only the inverse of a $p\times p$ matrix.
\section{Confidence Bands}\label{sec:confidence}
In Section~\ref{sec:bestPredictor} we suggested the
estimator $\Xhat$ for the continuation of the function
$\Xa$. In this section we would like to construct
confidence bands for this estimator. We consider two kinds
of confidence bands. The first is a global confidence band.
A global confidence band with confidence level
$(1-\delta)100\%$ is defined as a pair of functions, the
upper band $f_U$ and the lower band $f_L$, such that
$P(f_L(t)<\Xb(t)<f_U(t)\textrm{\;for\;all\;}t\in S_2)\geq
1-\delta$. We also consider local confidence bands. Local
confidence bands do not require that the last condition
holds simultaneously for all $t$; rather we are looking for
a pair of functions $g_U$ and $g_L$ such that for all $t\in
S_2$, $P(g_L(t)<\Xb(t)<g_U(t) )\geq 1-\delta$.

Our construction of both global and local confidence bands
is based on the technique introduced
by~\citet{KnaflSacksYlvisaker_band1985}. The idea is the
following. We first create simultaneous confidence
intervals for some finite set of point. Then, using the
attributes of spline functions, we complete this band for
all points of $S_2$. The computation of these bands can be
computationally demanding. Hence, we suggest also
confidence bands that are based on cross-validation. While
these confidence bands do not have the theoretical
guarantee of the former, they are simple to compute and
seem to work reasonably well (see
Section~\ref{sec:numerical}, Table~\ref{tb:confidence}).

In the following, we assume that $X$ and $Y$ are Gaussian
processes. Therefore $\Xb$ is also a Gaussian process and,
by~\eqref{eq:Gaussian},  $E[\Xb|\Ya]=\Xhat$. Similarly, we
have
\begin{equation*}
\Cov(\Xb(s),\Xb(t)|\Ya)=\bvecb(s)'\left(g_{22} -G_{21}\Gai G_{12}
\right)\bvecb(t)\,.
\end{equation*}
Define
\begin{equation*}
    Z(t)=\frac{\Xb(t)-\Xhat(t)}{\Var(\Xb(t)|\Ya)^{1/2}}\,,
\end{equation*}
then $Z(t)$ is a zero-mean Gaussian process with variance
$1$ for each $t$.

Let $t_1,\ldots,t_m$ be the breaks in $\tau_2$, i.e., the
knots of $\tau_2$, ignoring knot multiplicity. Let
$t_{i,j}=t_i+\frac{j-1}{k-1}(t_{i+1}-t_i)$,
$j=1,\ldots,k-1$. Define the following grid
\begin{equation*}
\mathcal{G}=\{t_{1,1},t_{1,2},\ldots,t_{m-1,k-1},t_m\}\,,
\end{equation*}
i.e., $\mathcal{G} $ is a grid that includes all the breaks
in $\tau_2$ and there are $k-2$ equally spaced grid points
between each two successive breaks of $\tau_2$. We are
interested in computing simultaneous confidence intervals
for the points in $\mathcal{G}$. In other words, for a
given $\delta$, we would like to find $z_{\delta}$ such
that
\begin{equation}\label{eq:z_delta}
P(\max_{t\in \mathcal{G}} |\Xb(t)-\Xhat(t)|>
z_{\delta}\Var(\Xb(t)|\Ya)^{1/2})=P(\max_{t\in \mathcal{G}} |Z(t)|>
z_{\delta})\leq \delta\,.
\end{equation}
$z_{\delta}$ can be found using simulations or by utilizing
the inequality \citep[][Eq.~(1.8)
]{KnaflSacksYlvisaker_band1985}
\begin{equation}\label{eq:maxGaussian}
    P(\max_{t\in \mathcal{G} } |Z(t)|> a )\leq
    P(|Z(t_{1,1})|>a)+\sum_{i=1}^{m-1}\sum_{j=1}^{k-1}P(|Z(t_{i,j})|\leq a\,,|Z(t_{i,j+1})|>a
    )\,.
\end{equation}

Recall that the trajectories of $(\Xb(t)-\Xhat)|Y_1$ are in
$ S_{k,\tau_2}$. Hence for each segment between two
successive breaks of $\tau_2$, say $[t_i,t_{i+1}]$, the
trajectories are $k$-ordered polynomials. Let $p(t)$ be a
restriction of such a trajectory to $[t_i,t_{i+1}]$. $p(t)$
can be written, using Lagrange polynomials, as
\begin{equation*}
p(t)=\sum_{j=1}^k \ell(t) p(t_{i,j})\quad ;\quad \ell_j(t) =
\prod_{r=1,\, r\neq j}^{k} \frac{t-t_{i,r}}{t_{i,j}-t_{i,r}}\,.
\end{equation*}
Note that for all $t\in [t_i,t_{i+1}]$, $|p(t)|\leq
\sum_{r=1}^k |\ell(t)| p(t_{i,j})$. Hence, if
\begin{equation}\label{eq:bound_point_polynomial}
      |p(t_{i,j})|<z_{\delta}\Var(\Xb(t_{i,j})|Y_1)^{1/2}) \textrm{\qquad for\,}j=1,\ldots,k\,,
\end{equation}
then for all $t\in[t_i,t_{i+1}]$
\begin{equation}\label{eq:bound_all_polynomial}
     |p(t)|< z_{\delta}\sum_{j=1}^k
    |\ell_j(t)|\Var(\Xb(t_{i,j})|Y_1)^{1/2})\doteq z_{\delta}D_{t_i}(t)\,.
\end{equation}
By~\eqref{eq:z_delta} we have that with probability greater
than or equal to $1-\delta$, the inequality
in~\eqref{eq:bound_point_polynomial} holds simultaneously
for all $i$. Thus, with probability greater than or equal
to $1-\delta$, the inequality
in~\eqref{eq:bound_all_polynomial} also holds. Define the
pair of functions $(f_U,f_L)$ on $S_2$ such that for all
$t\in [t_i,t_{i+1}]$
\begin{equation}\label{eq:globalCI}
f_U(t)=\Xhat(t)+ z_{\delta}D_{t_i}(t)\quad ;\quad  f_L(t)=\Xhat(t)-
z_{\delta}D_{t_i}(t) \,.
\end{equation}
Then $(f_U,f_L)$ are global confidence band for $\Xb|Y_1$
with a confidence level greater than or equal to
$100(1-\delta)\%$. Note that $f_U$ and $f_L$ are
continuous.

For local confidence bands, we can define the pair of
functions $(g_U,g_L)$ on $S_2$ such that for all $t\in
[t_i,t_{i+1}]$
\begin{equation}\label{eq:localCI}
g_U(t)=\Xhat(t)+ \hat{z}_{\delta}D_{t_i}(t)\quad ;\quad
g_L(t)=\Xhat(t)- \hat{z}_{\delta}D_{t_i}(t) \,,
\end{equation}
where
\begin{equation*}
\hat{z}_{\delta}=\max_{i}\min \left\{z_\delta :
P\left(\max_{j=1,\ldots,k } |Z(t_{i,j})|> z_{\delta}\right)\leq
\delta\right\}\,.
\end{equation*}
Using $\hat{z}_{\delta}$ ensures that $g_U$ and $g_L$ are
continuous. The estimation of $\hat{z}_{\delta}$ can be
done using the relation in~\eqref{eq:maxGaussian}. We note
that in the computation of $\hat{z}_{\delta}$ we demanded
that between each two successive breaks in $\tau_2$, with
probability greater than $1-\delta$ the trajectories of
$\Xb$ will stay within the band. While this can be
restrictive if the distance between successive points in
$\tau_2$ is large, a simple solution is to take the set
$\mathcal{G}$ to be more dense.

We remark here on some issues related to the confidence
bands defined in~(\ref{eq:globalCI}-\ref{eq:localCI}).
First, note that the bands are conservative, meaning that
the confidence level is greater than $100(1-\delta)\%$.
Second, we have assumed that $\Xb|Y_1$ is a Gaussian
process with known distribution. Third, the computation of
$z_{\delta}$ (or $\hat{z}_{\delta}$) can be demanding.
Hence, we suggest to estimate confidence bands from the
data using some sort of cross-validation. Compute
$\Var(\Xb(t)|\Ya)^{1/2} $ for all $t\in \mathcal{G}$, and
let $\hat{D}(t)$ be the $k$-ordered regression spline
function with knot sequence $\tau_2$ of the points
$\{(t,\Var(\Xb(t)|\Ya)^{1/2}) : t\in\mathcal{G}\}$. We
suggest the following confidence bands
\begin{equation}\label{eq:localCI_crossvalidation}
\hat{f}_U(t)=\Xhat(t)+  C_{\textrm{Global}} \hat{D}(t)\quad ;\quad
\hat{g}_U(t)=\Xhat(t)+ C_{\textrm{Local}} \hat{D}(t) \,,
\end{equation}
and similarly for $\hat{f}_L$ and $\hat{g}_L$ where
$C_{\textrm{Global}}$ and $C_{\textrm{Local}}$ are computed
using cross-validation as described below. Assume that the
functions $Y^{(1)},\ldots,Y^{(m)}$ were observed. Partition
the functions to $K$ folds ${F_j: j=1,\ldots,K}$. Compute
$\Xhat(t)$ and $\hat{D}(t)$ for each subset of $K-1$ folds.
Define
\begin{eqnarray*}
C_{\textrm{Global},j}&=&\min \left\{c>0 : \frac{1}{|F_j|}\sum_{Y_i\in
F_j} I\{|Y_i(t)-\Xhat(t)|<c
\hat{D}(t)\textrm{\;for\,all}\,t\in\mathcal{G}\}>1-\delta\right\}\\
C_{\textrm{Local},j}&=&\min \left\{c>0 :
\min_{t\in\mathcal{G}}\left(\frac{1}{|F_j|}\sum_{Y_i\in F_j}
I\{|Y_i(t)-\Xhat(t)|<c\hat{D}(t)\}\right)>1-\delta\right\}
\end{eqnarray*}
where $I\{B\}$ is the indicator function of the set $B$.
Then we suggest to choose $C_{\textrm{Global}}$ and
$C_{\textrm{Local}}$ to be the median of $
C_{\textrm{Global},j}$ and $C_{\textrm{Local},j}$
respectively. We note that the suggestion to extend the
confidence bands from points in the grid to the whole
segment using regression splines seems reasonable when the
grid is fine enough. In the numerical examples of
Section~\ref{sec:numerical} we compute the confidence bands
using the cross-validation technique.
\section{Numerical Examples}\label{sec:numerical}
In this section we apply the estimator $\Xhat$ to call
center data. We are interested in forecasting the
continuation of two processes: the arrival process and the
workload process. The estimators of these two processes
play an important roll in determining staffing level at
call centers \citep[see, for
example,][]{Aldor_2009,ShenHuang_2008, Michael_Thesis}.
Usually, staffing levels are determined in advance, at
least one day ahead. Here we propose a method for updating
the staffing level, given information obtained throughout
the beginning of the day. As noted by \citet{Gans2003} and
by \citet{ShenHuang_2008}, such updating is operationally
beneficial and feasible. If performed appropriately, it
could result in higher efficiency and service quality:
based on the revised forecasts, a manager can adjust
staffing levels correspondingly, by offering overtime to
agents on duty or dismissing agents early, calling in
additional agents if needed, increasing or reducing
cross-selling, and transferring agents to other activities
such as email inquiries and faxes.

This section is organized as follows. We first describe the
arrival and workload processes
(Section~\ref{sbsec:arrival_workload}). We then describe
the data (Section~\ref{sbsec:data}) and the forecast
implementation (Section~\ref{sbsec:implementation}). The
analysis appears in
Sections~\ref{sbsec:first}-\ref{sbsec:third}. Finally,
confidence bands are discussed in Section~\ref{sbsec:cb}.

\subsection{The arrival and workload processes}\label{sbsec:arrival_workload}

We define the arrival process of day $j$, $a_j(t)$, as the
number of calls that arrive on day $j$ during the time
interval $[t-c,t]$, where $t$ varies continuously over time
and $c$ is some fixed constant. Note that $a_j(t)$ itself
is not a continuous function, but when the call volume is
large and this function does not change drastically over
short time intervals, it can be assumed that the function
$a_j(t)$, for each day $j$, arises from some underlying
deterministic smooth arrival rate function $\lambda(t)$
plus some noise \citep{Weinberg2007}. In this case
$a_j(t)/c$ can be considered as an approximation of the
smooth function $\lambda(t)$. We now describe the workload
process $w_j(t)$ for each day $j$.  The function $w_j(t)$
counts the number of calls that would have been handled by
the call center on day $j$ at time $t$, \emph{assuming an
unlimited number of agents and hence no abandonments}. From
a management point of view, the advantage of looking at
$w_j(t)$ over looking at $a_j(t)$ is that $w_j(t)$ reflects
the number of agents actually needed at each point in time.
However, as opposed to the process $a_j(t)$, which is
observable in real time, the computation of $w_j(t)$, for a
specific time $t$, involves estimation of call durations
for abandoned calls and can be performed only after all
calls entered up to time $t$ are actuality served
\citep[see the discussion at][]{Aldor_2009,Michael_Thesis}.

\subsection{The data}\label{sbsec:data}
\begin{figure}[!ht]
\vskip 0.2in
\begin{center}
\includegraphics[width=5.2in]{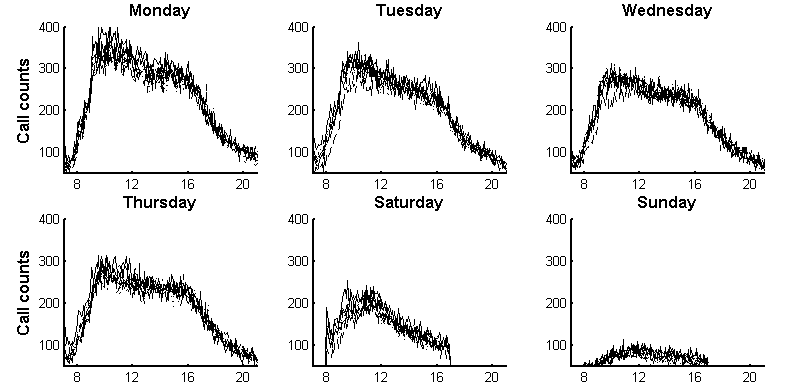}
\caption{Arrival count in five-minutes resolution for six successive
weeks, grouped according to weekday (Friday was omitted due to space
constraints). There is a clear difference between workdays,
Saturdays, and Sundays. For the working days, it seems that there is
some common pattern. Between 7~AM and 10~AM the call count rises
sharply to its peak. Then it decreases gradually until 4~PM. From
4~PM to 5~PM there is a rapid decrease followed by a more gradual
decrease from 5~PM until 12~AM. The call counts are smaller for
Saturday and much smaller for Sunday. Note also that the main
activity hours for weekends are 8~AM to 5~PM, as expected.
}\label{fig:days_of_the_week}
\end{center}
\vskip -0.2in
\end{figure}

The data used for the forecasting examples were gathered at
a call center of a large U.S. commercial bank. The bank has
various types of operations such as retail banking,
consumer lending, and private banking. Since the call
arrival pattern varies over different types of services, we
restrict attention to retail services, which account for
approximately 70\% of the calls
\citep[see][]{Weinberg2007}. The first two examples are of
the arrival process and the workload process, for weekdays
between March and October 2003. The data for the first
example consists of the arrival counts at five-minutes
resolution between 7:00~AM and 9:05~PM (i.e., $c=5$ in the
definition of $a_j(t)$). The data for the second example
consists of average workload, also in five-minutes
resolution, between 7:00~AM and 9:05~PM. There are 164 days
in the data set after excluding some abnormal days such as
holidays. Figure~\ref{fig:days_of_the_week} shows arrival
count profiles for different days of the week.

The third example explores the arrival process during
weekends between March and October 2003. There are 67 days
in the data set (excluding one day with incomplete data).
As can be seen from Figure~\ref{fig:days_of_the_week}, the
weekend behavior is different from that of the working
days, and there is a Saturday pattern and a Sunday pattern.
The data for this example consists of the arrival counts at
fifteen-minutes resolution between 8~AM and 5~PM. The
change in interval length from the previous two examples is
due to the decreased call-counts. The change in day length
is due to the low activity in early morning and late
afternoon hours on weekends (see
Figure~\ref{fig:days_of_the_week}).

In the first and second examples, we used the first 100
weekdays as the training set and the last 64 weekdays as
the test set. For each day from day 101 to day 164, we
extracted the \emph{same-weekday} information from the
preceding 100 days. Thus, for each day of the week we have
about 20 training days. For the third example, the test set
consists of weekend days 41 to 67 while the training set
for each day consist of its previous 40 weekend days. Thus,
similarly, for each day we have about 20 training days.
Additionally, we used the data from day start, up to 10~AM
and up to 12~PM. All forecasts were evaluated using the
data after 12~PM, which enabled fair comparison between the
results of the different cut points (10~AM and 12~PM). We
also compare our results to the mean of the preceding days,
from 12~PM on.

For a detailed description of the first example's data, the
reader is referred to \citet{Weinberg2007}, Section~2.  For
an explanation of how the second example's workload process
was computed, the reader is referred to
\citet{Michael_Thesis}. The data for the third example was
extracted using SEEStat, which is a software written at the
Technion SEELab\footnote{SEELab: The Technion Laboratory
for Service Enterprise Engineering. Webpage:
\url{http://ie.technion.ac.il/Labs/Serveng}\label{fn:seestat}}.
We refer the reader to \citet{USBANK2006} for a detailed
description of the U.S. commercial bank call-center data
from which the data for all three examples was extracted.
The U.S. bank call-center data is publicly downloaded from
SEESLab server\footnotemark[\value{footnote}].

\subsection{Forecast implementation}\label{sbsec:implementation}
The forecast was performed by Matlab implementation of the
BLUP algorithm from Section~\ref{sec:bestPredictor}, where
we enable regularization as in~\eqref{eq:ridge}. For the
implementation we used the functional data analysis Matlab
library, written by \citeauthor{bookRamsay}\footnote{The
functional data analysis Matlab library can be download
form
\url{ftp://ego.psych.mcgill.ca/pub/ramsay/FDAfuns/Matlab/}}.
The Matlab code, as well as the data sets, are downloadable
(see \ref{suppA}). The parameters for the forecast were
chosen using $10$-fold cross-validation (see end of
Section~\ref{sec:notation}). We computed local confidence
bands with $95\%$ confidence level using cross-validation,
as described in~\eqref{eq:localCI_crossvalidation}. We
quantified the results using both Root Mean Squared Error
(RMSE) and Average Percent Error (APE), which are defined
as follows. For each day $j$, let
\begin{equation*}
    RMSE_j=\left(\frac{1}{K}\sum_{k=1}^K
    (N_{jk}-\hat{N}_{jk})^2\right)^{1/2}\quad ; \quad
    APE_j=\frac{100}{K}\sum_{k=1}^K\frac{|
    N_{jk}-\hat{N}_{jk}|}{N_{jk}}\,,
\end{equation*}
where $N_{jk}$ is the actual number of calls (mean
workload) at the $k$-th time interval of day $j$ in the
arrival (workload) process application, $\hat{N}_{jk}$ is
the forecast of $N_{jk}$, and $K$ is the number of
intervals.

\subsection{First example: Arrival process for weekdays data}\label{sbsec:first}
Forecasting the arrival process for the first example data
was studied by both \citet{Weinberg2007} and
\citet{ShenHuang_2008}. \citeauthor{Weinberg2007} assumed
that the day patterns behave according to an autoregressive
model. The algorithm they suggest first gives a forecast
for the current day based on previous days' data. The
algorithm estimates the parameters in the autoregressive
model using Bayesian techniques. An update for the
continuation of the current day forecast is obtained by
conditioning on the data of the current day up to the cut
point. We refer to this algorithm as Bayesian update (BU)
for short. Similarly, the algorithm by
\citeauthor{ShenHuang_2008} assumes an autoregressive model
and gives a forecast for the current day. They then update
this forecast using least-square penalization, assuming an
underlying discrete process. We will refer to this
algorithm as penalized least square (PLS).

Comparison between the results of all three algorithms for
the first data set appears in Table~\ref{tb:comparison}.
Note that for all of the algorithms and all of the
categories there is improvement in the 10~AM and 12~PM
forecasts over the forecast based solely on past days. The
RMSE mean decreases by about 5-13\% for the 10~AM forecast,
and by 12-15\% for the 12~PM forecast, depending on the
algorithm. It should be noted that the algorithms by
\citeauthor{Weinberg2007} and by
\citeauthor{ShenHuang_2008} use information from all $100$
previous days and the knowledge of the previous day call
counts. In comparison, the BLUP algorithm uses only the
same weekday information ($\sim$20 days) and the previous
day information is not part of its training set.
Nevertheless, the results are similar.

\begin{table}[!b]
\begin{tabular}{c|ccccccccc}
\textbf{Example 1}& Previous day &  & \multicolumn{3}{c}{10 AM} &  & \multicolumn{3}{c}{12 PM}\\
 \cline{4-6}\cline{8-10}
 RMSE& mean &  & BU  &  PLS    &  BLUP &  & BU  &  PLS    &  BLUP\\
\hline
Minimum & 12.46 &  & \textbf{11.08} &  & 11.51 &  & \textbf{11.07} &  & 11.51\\
Q1 & 14.11 &  & 14.00 & \textbf{13.31} & 13.51 &  & 13.56 & 13.33 & \textbf{13.27}\\
Median & 16.40 &  & 15.50 & 14.87 & \textbf{14.69} &  & 14.80 & 14.60 & \textbf{14.17}\\
Mean & 19.11 &  & 17.86 & \textbf{16.48} & 16.83 &  & 16.59 & \textbf{16.13} & 16.15\\
Q3 & 21.27 &  & 19.87 & 17.26 & \textbf{17.04} &  & 16.58 & 16.39 & \textbf{15.92}\\
Maximum & 68.93 &  & 57.72 &  & \textbf{52.09} &  & 53.66 &  & \textbf{51.03}\\
\hline
\end{tabular}
 \vskip 0.2in \caption{
Summary of statistics (minimum, lower quartile (Q1),
median, mean, upper quartile (Q3), maximum) of RMSE for the
forecast based on the mean of the previous days, and BU,
PLS, and BLUP using data up to 10~AM and up to 12 PM for
the call arrival data set. The results for BU and PLS were
taken from the original papers. No maximum and minimum
results were given for PLS.
 }
 \label{tb:comparison}
\end{table}

\begin{figure}[!b]
\vskip 0.2in
\begin{center}
\includegraphics[width=5.5in]{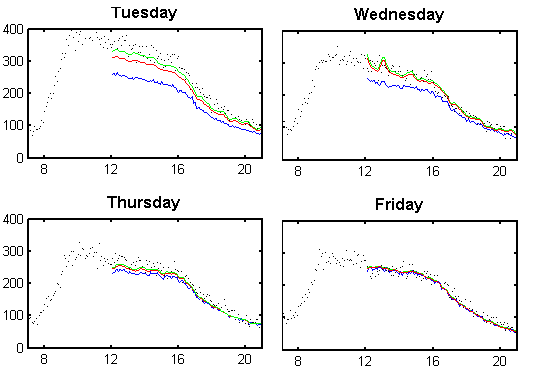}
\caption{Forecasting results for the week following Labor Day (Sept.
2-5, 2003) for the call arrival process of the first example. Labor
Day itself (Monday) does not appear since holiday data is not
included in the data set. The black dots represent the true call
counts in five-minutes resolution. The forecasts based on previous
days, 10~AM data, and 12~PM data are represented by the blue, red,
and green lines, respectively.}\label{fig:forecast}
\end{center}
\vskip -0.2in
\end{figure}

The forecasting results for the week that follows Labor Day
appear in Figure~\ref{fig:forecast}. It can be seen that
for the Tuesday that follows Labor Day (Monday) the call
counts are much higher than usual. This is captured, to
some degree, by the 10~AM forecast and much better by the
12~PM forecast. The same phenomenon occurs, with less
strength, during the Wednesday and Thursday following Labor
Day, until on Friday all the forecasts become roughly the
same. It seems that the power of the continuation-of-curve
forecasting is exactly in such situations, in which the
call counts are substantially different than usual
throughout the day, due to either predictable events, such
as holidays, or unpredictable events.

\subsection{Second example: Workload process for weekdays data}\label{sbsec:second}

The second example consists of the workload process for
weekdays data for the same period as the first example. We
forecast the workload process based on these sets of data:
previous days' data, up to 10~AM data, and up to 12 PM
data. We refer to this forecast as \em{direct} workload
forecast since we use past workload estimation as the basis
for the forecast. An alternative (and simpler) workload
forecasting method was proposed by \citet{Aldor_2009}.
\citeauthor{Aldor_2009} suggest to forecast the workload by
multiplying the forecasted arrival rate by the estimated
average service time \citep[see][Eq. 21]{Aldor_2009}. We
refer to this method as \em{indirect} workload forecasting.

\begin{figure}[!ht]
\vskip 0.2in
\begin{center}
\includegraphics[width=4.2in]{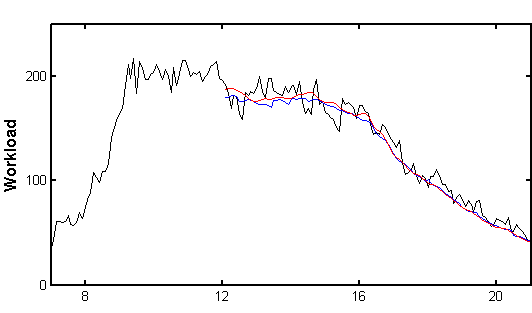}
\caption{Workload forecasting for Friday, September 5, 2003, using both the direct and the indirect methods. The black curve
represents the workload process estimated after observing the data
gathered throughout the day. The blue and red curves represents the
workload forecasts for the indirect and direct forecasts,
respectively, given data up to 12 PM.}\label{fig:comparison}
\end{center}
\vskip -0.2in
\end{figure}

Comparison between the two methods appears in
Table~\ref{tb:workload}. Following \citet{Aldor_2009}, we
estimated the average service time over a 30 minute period
for indirect workload computations. Note that the direct
workload forecast results are slightly better than the
indirect workload forecast in most of the categories. Also
note that in almost all categories, there is an improvement
in the 10~AM and 12~PM forecasts over the forecast based
solely on past days. The RMSE mean decreases by about 11\%
(9\%) for the 10~AM forecast, and by 15\% (12\%) for the
12~PM forecast for the direct (indirect) forecast.
Figure~\ref{fig:comparison} presents a visual comparison
between the direct and the indirect forecast methods on a
specific day. The two forecasts look roughly the same,
which is also true for all other days in this data set.

While in this example there is no significant difference
between the direct and indirect workload forecasts, we
expect these methods to obtain different forecasts when the
arrival rate changes during an average service time. This
is true, for example, for arrival and service of patients
in emergency rooms. The arrival rates of patients to
emergency rooms can change within an hour while the time
that a patient spends in emergency room (the ``service
time") is typically on the order of hours. As pointed out
by \citet[][Section 6]{Rozenshmidt2008}, in such cases,
forecasting the workload by the arrival count multiplied by
the average service time may not be accurate. This is
because the number of customers in the system is
cumulative, while the arrival count counts only those who
arrive in the current time interval. Thus, if the arrival
count is lower than it was in the previous time interval
and the average service time is long, the workload is
underestimated. Similarly, if the arrival count is larger
than previously, the workload is overestimated.
\begin{table}[!b]

\begin{tabular}{c|cccccccc}

\textbf{Example 2} & \multicolumn{2}{c}{Day ahead} &  & \multicolumn{2}{c}{10 AM} &  & \multicolumn{2}{c}{12 PM}  \\
\cline{2-3}\cline{5-6}\cline{8-9}
RMSE & Workload & Workload &  & Workload & Workload &  & Workload & Workload \\
 & (indirect) & (direct) &  & (indirect) & (direct) &  & (indirect) & (direct)  \\
\hline
Minimum & 8.72 & 8.41 &  & 7.98 & \textbf{7.71} &  & 7.96 & 8.50 \\
Q1 & 10.76 & 10.58 &  & 10.21 & 10.27 &  & 10.21 & \textbf{10.11} \\
Median & 12.10 & 12.26 &  & 11.63 & 11.21 &  & 11.66 & \textbf{11.05} \\
Mean & 15.97 & 15.95 &  & 14.59 & 14.26 &  & 14.13 & \textbf{13.48} \\
Q3 & 15.08 & 15.21 &  & 14.53 & 14.20 &  & 13.89 & \textbf{12.76} \\
Maximum & 96.09 & 94.79 &  & 95.74 & 85.11 &  & 93.39 & \textbf{71.20} \\
\end{tabular}
\vskip 0.2in \caption{ Summary of statistics (minimum,
lower quartile (Q1), median, mean, upper quartile (Q3),
maximum) of RMSE for the forecast based on the mean of the
previous days' data, up to 10~AM data and up to 12 PM data,
for the workload data set, for both the indirect and the
direct forecast methods using the BLUP.}\label{tb:workload}

\end{table}

\subsection{Third example: Arrival process for weekends data}\label{sbsec:third}
The third example it that of the weekends' arrivals. The
main difference between the first two examples and this one
is that the data in this example cannot be considered as
data from successive days, due the six day difference
between any Sunday and it successive Saturday. Recall that
the models considered by \citet{Weinberg2007} and
\citet{ShenHuang_2008} have an autoregressive structure.
Since this autoregressive structure seems not to hold for
this data, the techniques by \citeauthor{Weinberg2007} and
\citeauthor{ShenHuang_2008} are not directly applicable.
But even when the autoregressive structure does not hold,
the results appearing in Table~\ref{tb:weekends} reveal
that forecasting for this data set is still beneficial.
Indeed, the RMSE (APE) mean decreases by about 5\% (2\%)
for the 10~AM forecast, and by 10\% (4\%) for the 12~PM
forecast. While these results are not as good as the
results in the previous examples, note that the curves in
this example begin an hour later and while the call counts
are lower during weekends, the arrival rate variance does
not change drastically (see
Figure~\ref{fig:days_of_the_week}).

\begin{table}[!b]

\begin{tabular}{c|ccccccc}
 \textbf{Example 3} & \multicolumn{3}{c}{RMSE} &  & \multicolumn{3}{c}{APE}\\
\cline{2-4}\cline{6-8}
 & Day ahead & 10 AM & 12 PM &  & Day ahead & 10 AM & 12 PM\\
\hline
Minimum & 3.66 & 3.62 & 3.92 &  & 4.47 & 4.33 & 4.60\\
Q1 & 5.37 & 5.63 & 5.05 &  & 5.57 & 5.41 & 5.64\\
Median & 6.80 & 7.01 & 6.87 &  & 6.71 & 6.84 & 6.31\\
Mean & 7.64 & 7.19 & 6.97 &  & 7.23 & 7.10 & 6.97\\
Q3 & 9.01 & 8.97 & 8.59 &  & 8.83 & 8.16 & 7.44\\
Maximum & 16.12 & 11.84 & 11.13 &  & 12.17 & 11.80 & 12.46\\
\hline
\end{tabular}
\vskip 0.2in \caption{Summary of statistics (minimum, lower
quartile (Q1), median, mean, upper quartile (Q3), maximum)
of RMSE and APE for the forecast based on the mean of the
previous days and the BLUP, using 10~AM and 12~PM cuts for
the weekends data set. } \label{tb:weekends}
\end{table}

\subsection{Confidence bands}\label{sbsec:cb}
\begin{figure}[!ht]
\vskip 0.2in
\begin{center}
\includegraphics[width=4in]{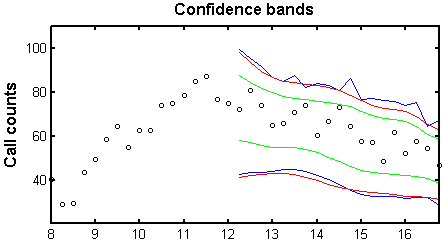}
\caption{Confidence bands for Sunday, August 10, 2003. The black
dots represent the true call counts in fifteen-minutes resolution.
The confidence bands based on previous days, 10~AM data, and 12~PM
data are represented by the blue, red, and green lines,
respectively.}\label{fig:band}
\end{center}
\vskip -0.2in
\end{figure}

Following \citet{Weinberg2007}, we introduce the 95\%
confidence band coverage (COVER) and the average 95\%
confidence band width (WIDTH). Specifically, for each day
$j$, let
\begin{equation*}
    COVER_j=\frac{1}{K}\sum_{k=1}^K
    I\left(F_{L,jk}<N_{jk}<F_{U,jk}\right)\; ; \;
    WIDTH_j=\frac{1}{K}\sum_{k=1}^K\left(F_{U,jk}-F_{L,jk}\right) \,,
\end{equation*}
where $(F_{L,jk},F_{U,jk})$ is the confidence band of day
$j$, evaluated at the beginning of the $k$-th interval (see
\eqref{eq:localCI_crossvalidation}). The mean coverage and
mean width, for all three examples, are presented in
Table~\ref{tb:confidence}. First, note that for all three
examples, the width of the confidence band narrows down as
more information is revealed. In other words, the width of
the confidence band for the 12~PM forecast is narrower than
the width for the 10~AM forecast which, in turn, is
narrower than the width for the pervious days' mean. We
also see that the mean coverage becomes more accurate as
more information is revealed. Figure~\ref{fig:band} depicts
the confidence bands for the arrival process on a
particular Sunday. Note that the confidence bands for the
previous days' forecast and the 10~AM forecast almost
coincide. However, at 12~PM, when enough information on
this particular day becomes available, the confidence band
narrows down and does capture the underlying behavior.

\begin{table}[!ht]
\begin{tabular}{cccccccc}
\hline
 & \multicolumn{3}{c}{ Coverage} &  & \multicolumn{3}{c}{ Width}\\
\cline{2-4}\cline{6-8}
 & Example 1 & Example 2 & Example 3 &  & Example 1 & Example 2 & Example 3\\
\hline
Mean & 93.19\% & 91.69\% & 98.15\% &  & 79.73 & 62.80 & 40.15\\
10 AM & 94.14\% & 92.27\% & 98.64\% &  & 74.99 & 56.45 & 39.53\\
12 PM & 94.86\% & 93.08\% & 96.49\% &  & 73.07 & 55.95 & 31.40\\
\hline
\end{tabular}
\vskip 0.2in
  \caption{The mean confidence band coverage and the mean width for the forecasts based on the previous days' mean, the 10~AM cut and the 12~PM cut
 for the arrival process on the working days data set (Example~1),
the workload process on the working days data set
(Example~2) and the arrival process on the weekends data
set (Example~3). }\label{tb:confidence}
\end{table}

Summarizing, using call center data, we demonstrated that
forecasting of curve continuation can be achieved
successfully by the proposed BLUP. We also showed that
confidence bands for such forecasts can be obtained using
cross-validation.
\section{Concluding Remarks}\label{sec:summary}

We have constructed the best linear unbiased predictor
(BLUP) for the continuation of a curve. We now add the
following comments regarding the construction of the BLUP
and its application to call center data.

First, in our analysis we have used a spline model to
describe the functions. This is not required for the
construction of the BLUP, and the proof of
Theorem~\ref{thm:hxisBLUP} holds for any function space of
finite dimension. However, as discussed previously, there
are two main advantages of using spline representation.
First, the computation of the covariance operators, for
$S_1, S_2$ and $S$ and between them, as well as the
pseudo-inverse covariance operator $\Gammai$, are all
computationally simple when working with splines. Second,
the representation of the restriction of a function to a
partial segment does not suffer from collinearity of the
basis functions, which can be the case for a more general
setting. Indeed, the number of basis elements can be
reduced significantly in the spline function model,
depending on the number of knots in the partial segment,
while the number of basis elements could remain the same in
a more general model.

Second, we have assumed that the random function $X$ lies
within a function space of (possibly large)
finite-dimension. While this is a restrictive assumption,
there are some difficulties with the BLUP definition (and
computation) for a random function that lies in an
infinite-dimensional space. The main difficulty is that
inverting the covariance operator (as done in
Lemma~\ref{lem:psaudoinverseWorks} for finite dimension) is
problematic since the inverse of the covariance operator
need not be bounded and may not exists. However, we believe
that characterization of the BLUP in the infinite-dimension
case is possible under some conditions. Further research is
required to address this question.

Finally, in this work we forecasted the continuation of the
workload process. As discussed in \citet{Feldman2008} and
\citet{Michael_Thesis}, the workload process is a more
appropriate candidate than the arrival process, as a basis
for determining staffing levels in call centers. This work,
along with \citet{Aldor_2009} and \citet{Michael_Thesis},
are the first steps in exploring direct forecasting of the
workload process, but more remains to done \citep[see, for
example][]{Whitt1999,Zeltyn2009}.
\appendix
\section{Proofs}
\subsection{Lemma~\ref{lem:psaudo_inverse}}
\begin{lem}\label{lem:psaudo_inverse}
Let $T$ be a $n\times p$ matrix of rank $s$ and let $L$ be
a $p\times p$ positive definite diagonal matrix. Then the
following assertions are true
\begin{enumerate}
  \item $T'T(T'T)^+ T'= T'$\label{lem:psaudo_inverse1}
  \item $T'LT(T'LT)^+ T'=
      T'$\label{lem:psaudo_inverse2}
  \item $(T'T)^{+}=T'\left((TT')^{+}\right)^2
      T$\label{lem:psaudo_inverse3}
\end{enumerate}
\end{lem}
\begin{proof}
\begin{enumerate}
  \item If $T'T$ is invertible then
      $(T'T)^+=(T'T)^{-1}$ and the result follows.
      Otherwise, let $U\Lambda V'$ be the
      svd~\citep[singular value decomposition,
      see][]{MatrixComputations} of $T$ where $U$ and
      $V$ are orthonormal columns matrices of size
      $n\times s$ and $p\times s$ respectively, and
      $\Lambda$ is a $s\times s$ positive definite
      diagonal matrix. Then
\begin{eqnarray*}
  T'T(T'T)^+ T'&=&(V\Lambda U')(U\Lambda V')\big((V\Lambda U')(U\Lambda V')\big)^+ V\Lambda U'\\
  & =& V \Lambda^2 V'(V \Lambda^2 V')^{+}V\Lambda U'\,.
\end{eqnarray*}
Since $\Lambda$ is invertible and $V$ has orthonormal
columns $(V \Lambda^2 V')^{+}=V\Lambda^{-2}V'$. Hence
\begin{equation*}
  T'T(T'T)^+ T'= V \Lambda^2 V'V \Lambda^{-2} V' V\Lambda U'= V\Lambda U'=T'\,.
\end{equation*}

  \item Denote  $W=L^{1/2}T$, then $T'LT(T'LT)^+
      T'=W'W(W'W)^+ W'L^{-1/2}$.
      Using~\ref{lem:psaudo_inverse1}., we obtain $ W'W
      (W'W)^+ W'L^{-1/2} = W'L^{-1/2}=T'$.
  \item Since $TT'$ is a positive semi-definite matrix,
      $TT'$ has a spectral representation of the form
      $TT'=\sum_{i=1}^s\lambda_i v_i v_i'$ where $s\leq
      \min\{n,p\}$, $\lambda_i>0$ and $\{v_i\}$ is an
      orthonormal set of vectors. Note that
      $TT'v_i=\lambda_i v_i$ and hence
      $T'T(T'v_i)=\lambda_i T'v_i$. Moreover
      $\norm{T'v_i}^2=v_i'TT'v_i=v_i'(\lambda_i
      v_i)=\lambda_i$. Hence, we obtained that
      $\left\{T'v_i/\sqrt{\lambda_i}\right\}$ is the
      set of orthonormal eigenvectors of $T'T$ with the
      respective non-zero eigenvalues $\{\lambda_i\}$.
      Thus,
\begin{equation*}
T'T=\sum_{i=1}^s
\lambda_i\frac{T'v_i}{\sqrt{\lambda_i}}\left(\frac{T'v_i}{\sqrt{\lambda_i}}\right)'=T'\left(\sum_{i=1}^s
v_i v_i' \right) T\,.
\end{equation*}
Using the spectral representation we also have
\begin{equation*}
    T'\left((TT')^{+}\right)^2T=T'\left(\sum_{i=1}^s\lambda_i^{-2}v_i
    v_i'\right)T\,.
\end{equation*}

In order to show that
$(T'T)^{+}=T'\left((TT')^{+}\right)^2 T$ we need to
show the following \citep[see][]{MatrixComputations}:
\begin{enumerate}
  \item $(T'T)\left(T'\left((TT')^{+}\right)^2
      T\right)(T'T)=(T'T)$
  \item $\left(T'\left((TT')^{+}\right)^2
      T\right)(T'T)(T'\left((TT')^{+}\right)^2
      T)=\left(T'\left((TT')^{+}\right)^2 T\right)$
  \item
      $\left((T'T)\left(T'\left((TT')^{+}\right)^2
      T\right)\right)'=(T'T)(T'\left((TT')^{+}\right)^2
      T)$
  \item $\left(\left(T'\left((TT')^{+}\right)^2
      T\right)(T'T)\right)'=(T'\left((TT')^{+}\right)^2
      T)(T'T)$
\end{enumerate}
 In order to see (a), note that
\begin{eqnarray*}
  (T'T)\left(T'\left((TT')^{+}\right)^2 T\right)(T'T)&=& T'( TT')\left(\sum_{i=1}^s\lambda_i^{-2}v_i
    v_i'\right)(TT') T \\
  &=&  T'\left(\sum_{i=1}^s \lambda_i v_i v_i' \right) \left(\sum_{i=1}^s\lambda_i^{-2}v_i
    v_i'\right)\left(\sum_{i=1}^s\lambda_i v_i v_i'\right)T\\
   &=&T'\left(\sum_{i=1}^s\ v_i v_i'\right)T=T'T\,.
\end{eqnarray*}
Similarly, for (b), we have
\begin{align*}
 \left(T'\left((TT')^{+}\right)^2 T\right)&(T'T)\left(T'\left((TT')^{+}\right)^2 T\right)=\\
 & =T'\left(\sum_{i=1}^s\lambda_i^{-2}v_i     v_i'\right)(TT')^2\left(\sum_{i=1}^s\lambda_i^{-2}v_i
    v_i'\right)T \\
  &=  T'\left(\sum_{i=1}^s \lambda_i^{-2} v_i v_i' \right) \left(\sum_{i=1}^s\lambda_i v_i
    v_i'\right)^2 \left(\sum_{i=1}^s\lambda_i^{-2} v_i v_i'\right)T\\
   &=T'\left(\sum_{i=1}^s\lambda_i^{-2} v_i v_i'\right)T=T'\left((TT')^{+}\right)^2T\,.
\end{align*}
For (c),
\begin{eqnarray*}
  \left((T'T)\left(T'\left((TT')^{+}\right)^2 T\right)\right)'&=& \left(T'( TT')\left(\sum_{i=1}^s\lambda_i^{-2}v_i
    v_i'\right)T\right)'   =  \left(T'\left(\sum_{i=1}^s \lambda_i^{-1} v_i v_i' \right)T\right)'\\
   &=&T'\left(\sum_{i=1}^s \lambda_i^{-1} v_i v_i' \right)T=  (T'T)\left(T'\left((TT')^{+}\right)^2 T\right)\,.
\end{eqnarray*}
Finally, (d) is shown similarly to (c).
\end{enumerate}\end{proof}

\subsection{Proof of Lemma~\ref{lem:psaudoinverseWorks}}
\begin{proof}
By~\eqref{not:vecY} we may write $\Ya(t)-\mu(t)=\bveca
(t)'(A_1 \hvec + B_1 \epsvec)$. Hence,
\begin{align*}
\left(  \Gamma_{11} \Gammai \right.&\left. (A_1 \hvec  + B_1
\epsvec)\right)(t)=\\  &= \bveca(t)' G_{11} W_1 W_1^{-1} \Gai (A_1
\hvec + B_1 \epsvec) \\&= \bveca(t)' G_{11} \Gai (A_1 \hvec + B_1
\epsvec)\\ &=\bveca(t)' [A_1,B_1]\left[
\begin{array}{cc}
  L&0\\0 &\Sigma
  \end{array}                                                                       \right]\left[
           \begin{array}{c}
          A_1 '\\
          B_1 '   \\
           \end{array}
         \right]\left([A_1,B_1]\left[                                                                      \begin{array}{cc}
  L&0\\0 &\Sigma
  \end{array}                                                                       \right]\left[
           \begin{array}{c}
          A_1 '\\
          B_1 '   \\
           \end{array}
         \right]\right)^+ \left[
           \begin{array}{c}
          A_1 \hvec\\
          B_1 \epsvec   \\
           \end{array}
         \right]
  \end{align*}
and the result follows from Lemma~\ref{lem:psaudo_inverse}.

Substituting $\hvec = L A_2'$ and $\epsvec=0$ in the last
equation, we also obtain
\begin{equation}\label{eq:GammaGammaInversegamm21}
G_{11} \Gai g_{12}= \Gamma_{11} \Gai (A_1 L A_2' + B_1
\textbf{0})=g_{12}\,.
\end{equation}
\end{proof}

\section*{Acknowledgements}
We thank Michael Reich for helpful discussions and for providing us
with the data for the workload example.

\begin{supplement}
\sname{Supplement A}\label{suppA} \stitle{Code and data
sets} \sdescription{Please read the file README.pdf for
details on the files in this folder.}
\slink[url]{http://pluto.huji.ac.il/~yaacov/blup.zip}
\end{supplement}

\bibliographystyle{imsart-nameyear}
%

\begin{thebibliography}{25}

\bibitem[\protect\citeauthoryear{Aldor-Noiman, Feigin and
  Mandelbaum}{2009}]{Aldor_2009}
\begin{bunpublished}[author]
\bauthor{\bsnm{Aldor-Noiman},~\bfnm{S.}\binits{S.}},
  \bauthor{\bsnm{Feigin},~\bfnm{P.~D.}\binits{P.~D.}} \AND
  \bauthor{\bsnm{Mandelbaum},~\bfnm{A.}\binits{A.}}
(\byear{2009}). \btitle{Workload forecasting for a call center:
\textsc{M}ethodology and a case
  study}.
\bnote{To appear}.
\end{bunpublished}
\endbibitem

\bibitem[\protect\citeauthoryear{Antoniadis, Paparoditis and
  Sapatinas}{2006}]{Antoniadis_autoregressive2006}
\begin{barticle}[author]
\bauthor{\bsnm{Antoniadis},~\bfnm{A.}\binits{A.}},
  \bauthor{\bsnm{Paparoditis},~\bfnm{E.}\binits{E.}} \AND
  \bauthor{\bsnm{Sapatinas},~\bfnm{T.}\binits{T.}}
(\byear{2006}). \btitle{A functional waveletkernel approach for time
series prediction}. \bjournal{Journal of the Royal Statistical
Society: Series B (Statistical
  Methodology)}
\bvolume{68} \bpages{837--857}.
\end{barticle}
\endbibitem

\bibitem[\protect\citeauthoryear{Besse, Cardot and
  Ferraty}{1997}]{BesseCardotFerraty97}
\begin{barticle}[author]
\bauthor{\bsnm{Besse},~\bfnm{P.}\binits{P.}},
  \bauthor{\bsnm{Cardot},~\bfnm{H.}\binits{H.}} \AND
  \bauthor{\bsnm{Ferraty},~\bfnm{F.}\binits{F.}}
(\byear{1997}). \btitle{Simultaneous non-parametric regressions of
unbalanced longitudinal
  data}.
\bjournal{Computational Statistics \& Data Analysis} \bvolume{24}
\bpages{255--270}.
\end{barticle}
\endbibitem

\bibitem[\protect\citeauthoryear{Besse, Cardot and
  Stephenson}{2000}]{BesseCardotStephenson_Autoregressive2000}
\begin{barticle}[author]
\bauthor{\bsnm{Besse},~\bfnm{Philippe~C.}\binits{P.~C.}},
  \bauthor{\bsnm{Cardot},~\bfnm{Herve}\binits{H.}} \AND
  \bauthor{\bsnm{Stephenson},~\bfnm{David~B.}\binits{D.~B.}}
(\byear{2000}). \btitle{Autoregressive forecasting of some
functional climatic variations}. \bjournal{Scandinavian Journal of
Statistics} \bvolume{27} \bpages{673--687}.
\end{barticle}
\endbibitem

\bibitem[\protect\citeauthoryear{de~Boor}{2001}]{bookDeBoor}
\begin{bbook}[author]
\bauthor{\bparticle{de~}\bsnm{Boor},~\bfnm{C.}\binits{C.}}
(\byear{2001}). \btitle{A practical guide to splines},
\bedition{Revised} ed. \bseries{Applied Mathematical Sciences}.
\bpublisher{Springer-Verlag New York}.
\end{bbook}
\endbibitem

\bibitem[\protect\citeauthoryear{Donin {\it et~al.}}{2006}]{USBANK2006}
\begin{bmisc}[author]
\bauthor{\bsnm{Donin},~\bfnm{O.}\binits{O.}},
  \bauthor{\bsnm{Feigin},~\bfnm{P.~D.}\binits{P.~D.}},
  \bauthor{\bsnm{Mandelbaum},~\bfnm{A.}\binits{A.}},
  \bauthor{\bsnm{Zeltyn},~\bfnm{S.}\binits{S.}},
  \bauthor{\bsnm{Trofimov},~\bfnm{V.}\binits{V.}},
  \bauthor{\bsnm{Ishay},~\bfnm{E.}\binits{E.}},
  \bauthor{\bsnm{Khudiakov},~\bfnm{P.}\binits{P.}} \AND
  \bauthor{\bsnm{Nadjharov},~\bfnm{E.}\binits{E.}}
(\byear{2006}). \btitle{The Call Center of "US Bank"}.
\bnote{Avaliable at
  \url{http://ie.technion.ac.il/Labs/Serveng/files/The_Call_Center_of_US_Bank.pdf}}.
\end{bmisc}
\endbibitem

\bibitem[\protect\citeauthoryear{Feldman {\it et~al.}}{2008}]{Feldman2008}
\begin{barticle}[author]
\bauthor{\bsnm{Feldman},~\bfnm{Z.}\binits{Z.}},
  \bauthor{\bsnm{Mandelbaum},~\bfnm{A.}\binits{A.}},
  \bauthor{\bsnm{Massey},~\bfnm{W.~A.}\binits{W.~A.}} \AND
  \bauthor{\bsnm{Whitt},~\bfnm{W.}\binits{W.}}
(\byear{2008}). \btitle{Staffing of Time-Varying Queues to Achieve
Time-Stable Performance}. \bjournal{Management Science} \bvolume{54}
\bpages{324--338}.
\end{barticle}
\endbibitem

\bibitem[\protect\citeauthoryear{Gans, Koole and Mandelbaum}{2003}]{Gans2003}
\begin{barticle}[author]
\bauthor{\bsnm{Gans},~\bfnm{N.}\binits{N.}},
  \bauthor{\bsnm{Koole},~\bfnm{G.}\binits{G.}} \AND
  \bauthor{\bsnm{Mandelbaum},~\bfnm{A.}\binits{A.}}
(\byear{2003}). \btitle{{Telephone Call Centers: Tutorial, Review,
and Research Prospects}}. \bjournal{Manufacturing Service Operations
Management} \bvolume{5} \bpages{79-141}.
\end{barticle}
\endbibitem

\bibitem[\protect\citeauthoryear{Golub and Loan}{1983}]{MatrixComputations}
\begin{bbook}[author]
\bauthor{\bsnm{Golub},~\bfnm{G.~H.}\binits{G.~H.}} \AND
  \bauthor{\bsnm{Loan},~\bfnm{C.~F.~Van}\binits{C.~F.~V.}}
(\byear{1983}). \btitle{Matrix computations}. \bpublisher{Johns
Hopkins University Press}, \baddress{Baltimore, Maryland}.
\end{bbook}
\endbibitem

\bibitem[\protect\citeauthoryear{Hoerl and Kennard}{1970}]{HoelKennard70}
\begin{barticle}[author]
\bauthor{\bsnm{Hoerl},~\bfnm{A.~E.}\binits{A.~E.}} \AND
  \bauthor{\bsnm{Kennard},~\bfnm{R.~W.}\binits{R.~W.}}
(\byear{1970}). \btitle{Ridge regression: biased estimation for
nonorthogonal problems}. \bjournal{Technometrics} \bvolume{12}
\bpages{55--67}.
\end{barticle}
\endbibitem

\bibitem[\protect\citeauthoryear{Knafl, Sacks and
  Ylvisaker}{1985}]{KnaflSacksYlvisaker_band1985}
\begin{barticle}[author]
\bauthor{\bsnm{Knafl},~\bfnm{G.}\binits{G.}},
  \bauthor{\bsnm{Sacks},~\bfnm{J.}\binits{J.}} \AND
  \bauthor{\bsnm{Ylvisaker},~\bfnm{D.}\binits{D.}}
(\byear{1985}). \btitle{Confidence bands for regression functions}.
\bjournal{Journal of the American Statistical Association}
\bvolume{80} \bpages{683--691}.
\end{barticle}
\endbibitem

\bibitem[\protect\citeauthoryear{Kneip}{1994}]{Kneip94}
\begin{barticle}[author]
\bauthor{\bsnm{Kneip},~\bfnm{A.}\binits{A.}} (\byear{1994}).
\btitle{Nonparametric estimation of common regressors for similar
curve data}. \bjournal{The Annals of Statistics} \bvolume{22}
\bpages{1386--1427}.
\end{barticle}
\endbibitem

\bibitem[\protect\citeauthoryear{Marsaglia}{1964}]{singularCovariance_Marsagli%
a64}
\begin{barticle}[author]
\bauthor{\bsnm{Marsaglia},~\bfnm{G.}\binits{G.}} (\byear{1964}).
\btitle{Conditional means and covariances of normal variables with
singular
  covariance matrix}.
\bjournal{Journal of the American Statistical Association}
\bvolume{59} \bpages{1203--1204}.
\end{barticle}
\endbibitem

\bibitem[\protect\citeauthoryear{Ramsay and
  Silverman}{2002}]{bookRamsayApplied}
\begin{bbook}[author]
\bauthor{\bsnm{Ramsay},~\bfnm{J.}\binits{J.}} \AND
  \bauthor{\bsnm{Silverman},~\bfnm{B.~W.}\binits{B.~W.}}
(\byear{2002}). \btitle{Applied functional data analysis: methods
and case studies}, \bedition{2nd} ed. \bseries{Springer Series in
Statistics}. \bpublisher{Springer-Verlag New York}.
\end{bbook}
\endbibitem

\bibitem[\protect\citeauthoryear{Ramsay and Silverman}{2005}]{bookRamsay}
\begin{bbook}[author]
\bauthor{\bsnm{Ramsay},~\bfnm{J.}\binits{J.}} \AND
  \bauthor{\bsnm{Silverman},~\bfnm{B.~W.}\binits{B.~W.}}
(\byear{2005}). \btitle{Functional data analysis}. \bseries{Springer
Series in Statistics}. \bpublisher{Springer-Verlag New York}.
\end{bbook}
\endbibitem

\bibitem[\protect\citeauthoryear{Reich}{2010}]{Michael_Thesis}
\begin{bmastersthesis}[author]
\bauthor{\bsnm{Reich},~\bfnm{M.}\binits{M.}} (\byear{2010}).
\btitle{The workload process: modelling, inference and applications}
\btype{Master's thesis}, \bschool{Technion - Israel Institute of
Technology.} \bnote{In preparation. The proposal is avaliable at
  \url{http://ie.technion.ac.il/serveng/References/references.html}}.
\end{bmastersthesis}
\endbibitem

\bibitem[\protect\citeauthoryear{Robinson}{1991}]{BLUP_Robinson91}
\begin{barticle}[author]
\bauthor{\bsnm{Robinson},~\bfnm{G.~K.}\binits{G.~K.}}
(\byear{1991}). \btitle{That \textsc{BLUP} is a good thing: the
estimation of random effects}. \bjournal{Statistical Science}
\bvolume{6} \bpages{15--32}.
\end{barticle}
\endbibitem

\bibitem[\protect\citeauthoryear{Rozenshmidt}{2008}]{Rozenshmidt2008}
\begin{bmastersthesis}[author]
\bauthor{\bsnm{Rozenshmidt},~\bfnm{L.}\binits{L.}} (\byear{2008}).
\btitle{On priority queues with impatient customers: Stationary and
  time-varying analysis}
\btype{Master's thesis}, \bschool{Technion - Israel Institute of
Technology.} \bnote{Avaliable at
  \url{http://iew3.technion.ac.il/serveng/References/thesis_Luba_Eng.pdf}}.
\end{bmastersthesis}
\endbibitem

\bibitem[\protect\citeauthoryear{Sansone}{1991}]{Sansone59}
\begin{bbook}[author]
\bauthor{\bsnm{Sansone},~\bfnm{G.}\binits{G.}} (\byear{1991}).
\btitle{Orthogonal functions}, \bedition{Rev.~ed.} ed.
\bpublisher{Dover Publications,}, \baddress{New York}.
\end{bbook}
\endbibitem

\bibitem[\protect\citeauthoryear{Shen}{2009}]{Shen_2009}
\begin{barticle}[author]
\bauthor{\bsnm{Shen},~\bfnm{H.}\binits{H.}} (\byear{2009}).
\btitle{On modeling and forecasting time series of smooth curves}.
\bjournal{Technometrics} \bvolume{51} \bpages{227--238}.
\end{barticle}
\endbibitem

\bibitem[\protect\citeauthoryear{Shen and Huang}{2008}]{ShenHuang_2008}
\begin{barticle}[author]
\bauthor{\bsnm{Shen},~\bfnm{H.}\binits{H.}} \AND
  \bauthor{\bsnm{Huang},~\bfnm{J.~Z.}\binits{J.~Z.}}
(\byear{2008}). \btitle{Interday Forecasting and Intraday Updating
of Call Center Arrivals}. \bjournal{Manufacturing Service Operations
Management} \bvolume{10} \bpages{391--410}.
\end{barticle}
\endbibitem

\bibitem[\protect\citeauthoryear{Weinberg, Brown and
  Stroud}{2007}]{Weinberg2007}
\begin{barticle}[author]
\bauthor{\bsnm{Weinberg},~\bfnm{J.}\binits{J.}},
  \bauthor{\bsnm{Brown},~\bfnm{L.~D.}\binits{L.~D.}} \AND
  \bauthor{\bsnm{Stroud},~\bfnm{J.~R.}\binits{J.~R.}}
(\byear{2007}). \btitle{Bayesian forecasting of an inhomogeneous
poissonprocess with
  applications to call center data}.
\bjournal{Journal of the American Statistical Association}
\bvolume{Vol. 102}.
\end{barticle}
\endbibitem

\bibitem[\protect\citeauthoryear{Whitt}{1999}]{Whitt1999}
\begin{barticle}[author]
\bauthor{\bsnm{Whitt},~\bfnm{W.}\binits{W.}} (\byear{1999}).
\btitle{Dynamic staffing in a telephone call center aiming to
immediately
  answer all calls}.
\bjournal{Operations Research Letters} \bvolume{24} \bpages{205 -
212}.
\end{barticle}
\endbibitem

\bibitem[\protect\citeauthoryear{Zeltyn}{2005}]{Zeltyn_2005}
\begin{bphdthesis}[author]
\bauthor{\bsnm{Zeltyn},~\bfnm{S.}\binits{S.}} (\byear{2005}).
\btitle{Call centers with impatient customers: \textsc{E}xact
analysis and
  many-server asymptotics of the \textsc{M}/\textsc{M}/n+\textsc{G} queue.}
\btype{PhD thesis}, \bschool{Technion—Israel Institute of
Technology.} \bnote{Available at
\url{http://ie.technion.ac.il/serveng/References/references.html}}.
\end{bphdthesis}
\endbibitem

\bibitem[\protect\citeauthoryear{Zeltyn {\it et~al.}}{2009}]{Zeltyn2009}
\begin{bunpublished}[author]
\bauthor{\bsnm{Zeltyn},~\bfnm{S.}\binits{S.}},
  \bauthor{\bsnm{Carmeli},~\bfnm{B.}\binits{B.}},
  \bauthor{\bsnm{Greenshpan},~\bfnm{O.}\binits{O.}},
  \bauthor{\bsnm{Mesika},~\bfnm{Y.}\binits{Y.}},
  \bauthor{\bsnm{Wasserkrug},~\bfnm{S.}\binits{S.}},
  \bauthor{\bsnm{Vortman},~\bfnm{P.}\binits{P.}},
  \bauthor{\bsnm{Marmor},~\bfnm{Y.~N.}\binits{Y.~N.}},
  \bauthor{\bsnm{Mandelbaum},~\bfnm{A.}\binits{A.}},
  \bauthor{\bsnm{Shtub},~\bfnm{A.}\binits{A.}},
  \bauthor{\bsnm{Lauterman},~\bfnm{T.}\binits{T.}},
  \bauthor{\bsnm{Schwartz},~\bfnm{D.}\binits{D.}},
  \bauthor{\bsnm{Moskovitch},~\bfnm{K.}\binits{K.}},
  \bauthor{\bsnm{Tzafrir},~\bfnm{S.}\binits{S.}} \AND
  \bauthor{\bsnm{Basis},~\bfnm{F.}\binits{F.}}
(\byear{2009}). \btitle{Simulation-Based Models of Emergency
Departments: Operational, Tactical
  and Strategic Staffing}.
\bnote{Under review}.
\end{bunpublished}
\endbibitem

\end{thebibliography}

\end{document}